# Asymmetric Diamond Emitters for Unidirectional Photon Coupling

SATYAJIT MURMU, AVIJIT KUMAR, AND RAJAN JHA *

[1]Nanophotonics and Plasmonics Laboratory, School of Basic Sciences, Indian Institute of Technology Bhubaneswar, Odisha 752050, India
*Corresponding author: *rjha@iitbbs.ac.in

The demand of single photon coupling to optical systems requires a suitable interface between negatively charged Nitrogen Vacancy (NV-) emitters in diamond to waveguides or optical circuits for applications in quantum network systems and in unidirectional emission with controlled photon states. Here, we propose a hybrid asymmetric structure of elliptically-faceted (ELFA) diamond nanowire with Bragg Grating containing negatively charged NV center for efficient and unidirectional optical coupling to optical nanowire. Our calculations indicate that the structure can provide coupling efficiency of ~90% towards the elliptically facet direction and ~1% towards the opposite direction. Further, Purcell factor is enhanced due to the augmented electric field intensity in the Bragg Grating assisted ELFA structure. Further, we observe higher chirality constant for ELFA+BG structure- an indicative of efficient unidirectional photon coupling. By integrating two ELFA diamond nanowire having opposite aligned elliptical facet can be used to design and develop complex optical circuits. Further, this structure can also be used as an inline polarizer filter where it can reflect only one particular mode (e.g. TE mode or TM mode) with higher extinct ratio. The hybrid structure can potentially be used for various applications in quantum photonics, quantum-nonlinear systems.

Nitrogen Vacancy (NV-) defect center in diamond is an example of quantum emitters possessing extraordinary optical and magnetic properties [1], such as coherent optical interface between electronic and nuclear spin [2, 3], narrow optical transition [4], and high temporal coherence of paramagnetic electron spin resonance at room temperature [5]. Further, it emits single photons with high photostability through both continuous and pulse excitation. The intrinsic single-photon emission properties of the emitters over a wide temperature range enable them for their wide-field applications in quantum memories [6], magnetometry [7], quantum computing [8], opto-spin based chemical sensing [9], or in general a single-photon source [10, 11]. Some optical systems such as photonic crystal cavities [12], micro cavities [13], solid immersion lenses [14] are used as interface for photon coupling. Further, processed single mode fiber (SMF) i.e. optical nanowire has also been used to collect single photon from NV- centers in diamond [15,16]. Interfacing of the quantum emitters of chiral molecules or quantum dots for unidirectional light coupling is of significant interest as it provides non-classical mapping of matter state into the optical state [17]. Generally, in systems, spin-momentum locking between spin of exciton states and angular momentum of light emission enables the spin states to control the propagation direction of light. Similarly, in plasmonic systems also an ideal unidirectional coupler has been designed with a circularly polarized dipole [18]. Further, the free space light has been unidirectionally guided into waveguides using sub-grating technique by implementation of the spin-momentum coupling concept [19]. The unidirectionality concept is needed to encode the information in photon states as qubits and also applied to communication, and complex optical circuits.

Here, we theoretically report a unidirectional light coupling of a linearly polarized dipole (NV- center) emitter embedded in a diamond nanowire coupled to an optical nanofiber. We propose an elliptically faceted (ELFA) diamond nanowire integrated with an optical nanowire in order to achieve enhanced unidirectional light coupling. The ELFA structure exhibits the unidirectional coupling and by incorporating a Bragg grating (BG) in the ELFA (ELFA+BG) structure further enhances the unidirectional coupling efficiency as the spontaneous emission rate of NV- center increases due to the presence of BG. The ELFA+BG structure shows a maximum unidirectional coupling efficiency of 90 % with significant Purcell factor. Integrating two ELFA diamond nanowire having opposite aligned elliptical facet can be used to design and develop complex optical circuits. Further, this structure can also be used as an inline polarizer filter where it can reflect only one particular mode (e.g. TE mode or TM mode) with higher extinct ratio. The hybrid structure can potentially be used for various applications in quantum photonics, quantum-nonlinear systems [20].

Figure 1(a) shows a three-dimensional schematic diagram of ELFA+ BG structure coupled with an optical nanowire. ELFA nanowire is a diamond nanowire containing an elliptical facet on one side and circular facet on the other side as shown in Fig. S1(a) in Supplementary Information (SI). The diamond nanorods can be realized using reactive ion etching and template processes [21] and the air grooves can be realized by using focused ion-beam process. Elliptical facet can experimentally be realized by controlled e-beam lithography [15]. Other alternative ways to achieve the desired etching angle in diamond nanowire is through focused ion-beam processing [22, 23]. The ELFA diamond nanowire has diameter $D_{nd}$, etched length of elliptical facet $L_e$, and free length (length of cylindrical part) as $L_f$ as shown in Fig. 1(b). The optical nanowire has diameter $D_{nf}$. The parallel integration of optical nanowire with ELFA diamond nanowire allows coupling of light which can vary from diabatic to the adiabatic process depending upon the value of $L_e$ [24]. We performed numerical simulations using FDTD for this structure by assuming refractive indices for optical and diamond nanowire as 1.450 and 2.410, respectively. Further, an oscillating dipole emitter has been placed at $(x, y, z) \equiv (0, 285, 0)$ to mimic the NV$^-$ defect center with photoemission at 637 nm as shown in Figure 1(b) (a red dot and arrow). The z-axis position of the dipole coincides with the onset of the elliptical facet ($z = 0$). Bragg's grating consists of azimuthally oriented air-grooves w.r.t optical nanowire which are parallel to the x-axis. While $L_d$ is the separation between the position of the dipole and the first air groove, a and b are the periodicity and the diameter of the air-grooves, respectively.

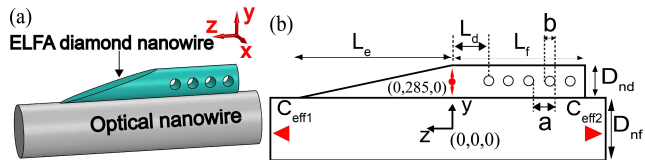

Figure 1. Proposed structure: (a) schematic of hybrid ELFA+BG structure coupled with optical nanowire. (b) 2D geometry of ELFA+BG structure at x=0 plane. The coupling efficiency along the z-direction (along elliptical facet direction) is denoted as $C_{eff1}$ and in the opposite it is $C_{eff2}$.

In semi-classical approach, the spontaneous emission rate is proportional to the available final density of states according to Fermi's golden rule. In case of coupling of dipole into waveguides these are known as local density of final photon states, which are different than free space density of states [25]. Thus, the dipole emission rate gets modified compared to free space excitation radiation due to the supported electric field within the waveguide system, cavity, and resonator. The quantity of this modified emission rate can be expressed as Purcell factor ($\Gamma$) and is given by the ratio of total spontaneous emission rate in the waveguide/cavity system to the free space spontaneous emission rate. In the proposed inhomogeneous system, the total emission rate can be given by the summation of the guided mode emission rate and radiative mode emission rate as in the ref.[26]. Then the coupling efficiency can be given by the ratio of guided mode emission rate to the total spontaneous emission rate of dipole. In order to further quantify unidirectional coupling efficiency of photon emission in our hybrid structure, we have defined the chirality constant as $(C_{eff1}-C_{eff2})/(C_{eff1}+C_{eff2})$ [27]. Here, $C_{eff1}$ and $C_{eff2}$ are the coupling efficiencies towards the elliptical facet direction and the opposite direction (circular facet direction), respectively, as shown in Figure 1(b). For an ideal unidirectional coupling, the chirality constant should be one.

A strong coupling of light from the supermode of the hybrid structure to optical nanowire fundamental mode can be described as the dipole emission first getting coupled to electric field of the supermode followed by coupling to optical nanowire through successive electric fields of intermediate cross-sections in elliptically facet region [see SI Fig. S1]. Here, the effective mode index values decreases and subsequently converge to optical nanowire fundamental mode towards the positive z-axis. One can see the existence of second-order supermode in ELFA diamond nanowire coupled to an optical nanowire structure as shown in SI Fig. S1. In the light coupling process, the adiabatic theorem plays a major role for efficient transfer of power through first-order supermode[24] as discussed later. Accordingly, $L_f$ and $L_e$ would play a major role in light coupling efficiency and can be tuned to achieve significantly high coupling efficiency.

Fig 2(a) shows normalized reflection spectra for both ELFA and ELFA + BG structure for each input light polarization. The spectra has been calculated for the optimized parameters $D_{nd} = 350$ nm, $D_{nf} = 220$ nm, $L_f = 2.7$ μm, $L_e = 6$ μm, $L_d = 80$ nm, a = 205 nm, b = 80 nm, and the BG containing 11 air grooves. It is evident that TM-like supermode (input y-polarized light) shows a very high reflectivity for bandwidth of 90 nm compared to the TE-like supermode (input x-polarized light) for the ELFA+BG structure. On the other hand, the reflectivity is very small for both the modes for the ELFA structure.

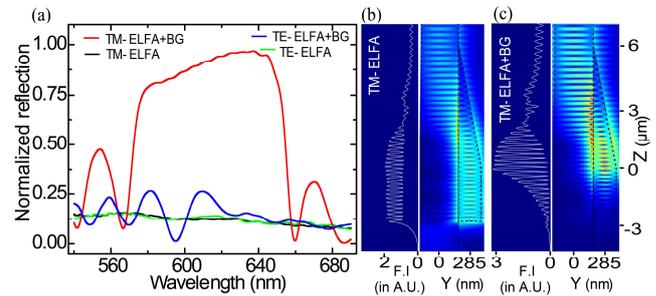

Figure 2. (a) Reflection spectra for ELFA and ELFA+BG hybrid structure with TM-polarized light and TE-polarized incident light through the elliptical facet side. Electric field profile (left panel) and distribution (right panel) of the reflected light in the x = 0 plane for (b) ELFA and (c) for ELFA+BG. Here, F.I. indicates field intensity.

Further evidence of difference in reflectivities can be inferred from electric field intensity plots in Fig. 2 (b) and 2 (c). The respective right panels correspond to the electric field intensity distribution in x=0 plane for ELFA structure and ELFA+BG structure coupled with optical nanowire. In the case of ELFA structure, the light gets reflected from the circular facet and the electric field largely remains similar throughout the cylindrical part of the diamond nanowire. This is due to the higher effective mode index of the ELFA and optical nanowire supermode than that of optical nanowire mode (see SI Fig. S1(b)). However, with the addition of BG, there is significant increase in the electric field intensity in the cylindrical region as can be seen from left panels of Fig. 2 (b)-(c). We notice that at z = 0, the electric field intensity shows the maximum value in the diamond nanowire for ELFA+BG structure i.e. the BG is perfectly matched to the ELFA structure with above mentioned geometrical parameters. As the normalized reflectivity for the TM-mode is significantly higher than TE-mode in ELFA+BG structure, this may be used to filter out the TE-mode. Further, by tuning the periodicity of the BG, one can make use of this concept in many nano-optical systems for other wavelengths.

To investigate the reflectivity profile and transmissivity profile from the EFLA structure for different structure parameters, at 637 nm wavelength of input light, we have calculated various parameters for different $L_f$ at fixed $L_e = 6$ μm as shown in Fig. 3. We observe periodicity in the reflectivity profile and the transmissivity profile with different periodicities. The reflection profile has two modulations with the small modulation period of 200 nm and the large modulation period of 1.4 μm. The observed modulations in the reflection profile may be attributed to the interference due to the multiple reflection of the first and second order supermodes from the circular facet. On the other hand, the transmission profile shows one modulation with modulation period of 1.4 μm which may be attributed to an interference between first and the second order supermodes only [See SI Fig. S2]. Further, we observe up and down spikes superimposed in both the transmission and the reflection profiles. The spikes have periodicity of 60 nm and indicate multibeam interference as a result of high order reflections from the circular and the elliptically facets. Fig. 3b shows cross-sectional field distribution at the location of the dipole (0, 285, 0) for $L_f$ = 10, 70, 100, 130, 190 and 240 nm. At $L_f$ = 70, 130, and 190 nm, we observe a step like feature in the electric field distribution along the y-axis (at x=0) which corresponds to the spikes in the reflection and transmission profiles. At the other values of $L_f$, we simply observe electric field distribution due to the interference between the various modes. The field distribution at other $L_f$ value are shown in SI Fig. S2(c). Placing a dipole emitter at the location of maximum electric field intensity will enhance the Purcell factor ($\Gamma$) significantly. Coupling efficiencies and the Purcell factor of ELFA structure has been described in SI Fig. S3.

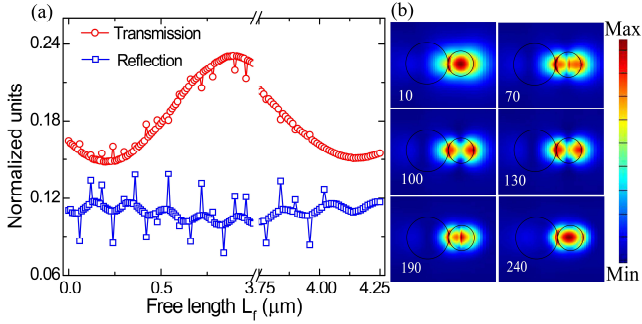

Figure 3. (a) Transmission and Reflection profiles at 637 nm for various $L_f$ for ELFA structure. (b) The field intensity distribution at dipole plane for $L_f$ = 10, 70, 100, 130, 190, and 240 nm.

Figure 4(a) shows $\Gamma$ variation with $L_e$ corresponding to y-polarized dipole emission for ELFA (blue curve) and ELFA+BG structures (red curve) with the same structural parameters as has been used for Fig. 2. We find out that $\Gamma$ is significantly higher for ELFA+BG as compared to that of ELFA structure due to higher electric field intensity as discussed earlier and which also further investigated. The spikes in $\Gamma$ values corresponds to period of 60 nm which is similar to that as observed in Fig. 3. Figure 4 (b), shows the coupling efficiency variation with $L_e$ for the two structures. We found that for both the structures, $C_{eff1}$ is significantly higher which corresponds to optical coupling along the elliptical facet direction and significantly smaller in the opposite direction ($C_{eff2}$) (see fig. 1(b)) thereby implying unidirectional coupling of the light. While ELFA structure provides a $C_{eff1}$ >50 %, ELFA+BG structure provides $C_{eff1}$ = 90 % which is the highest to the best of our knowledge. However, the $C_{eff2}$ remains less than 1% irrespective of the value of $L_e$. Further, $C_{eff1}$ increases with $L_e$ and nearly saturates at 2 μm beyond which there is an oscillatory behavior with a period ~4 μm for both ELFA and ELFA+BG structure. In SI Fig. S6, we estimate unidirectional contrast (Chirality Constant) as function of $L_e$ which tends to saturate at $L_e$ = 2 μm for both structures. The Chirality constant for ELFA+BG is higher than 0.90 and it's saturating to 0.98 for larger value of $L_e$. This clearly suggests the capability of the proposed structure for unidirectional photon coupling to the optical nanowire. We have calculated coupling efficiencies $C_{eff1}$ and $C_{eff2}$ and Purcell factor for various $L_d$ as shown in Figure 4(c). Here we observe oscillatory type behavior as function of $L_d$. At $L_d$ = 0 and $L_d$ ~180 nm, the $C_{eff1}$ and Purcell factor is very small, but the $C_{eff2}$ increases slightly. Similarly for the value of $L_d$ ~ 80 and ~280 nm, the $C_{eff1}$ is nearly 90% and $C_{eff2}$ is very small. This corresponds to Purcell factor equal to 6.2. This implies that the light reflected from the circular facet and BG should be in phase to achieve efficient unidirectional light coupling which also results in higher Purcell factor. Further, Fig. 4(d) shows that the coupling efficiency and Purcell factor for various number of grating period (N) in ELFA+BG structure. The $C_{eff1}$ and $\Gamma$ increase with N and tend to saturate for N>6 due to enhanced reflectivity with increasing N. On the other hand, $C_{eff2}$ decreases to even lower values with increasing N. These results may have implication in changing the photon state and also the values of second order correlation function with significant potential application in quantum optical circuits.

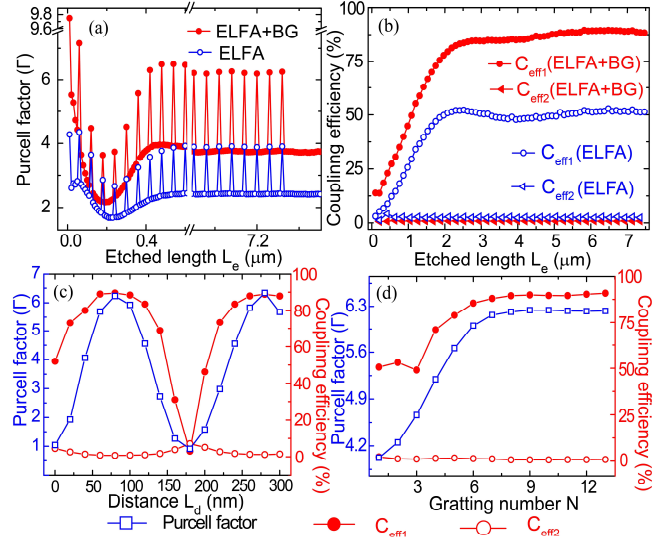

Figure 4. (a) Purcell factor for different etched length with constant other structure parameter as earlier discussed. (b) Coupling efficiency for same as Fig (a) but $L_e$ = 0.1 to $L_e$ = 7.5 μm. (c) Variation of $\Gamma$, $C_{eff1}$, and $C_{eff2}$ with different dipole position on axial line along circular facet ($L_f$ = 2.7 μm). (d) Variation of coupling efficiencies and Purcell factor w.r.t the grating no (N) in ELFA+BG structure. For Fig. (c) and (d) the common legend are given at lower portion of Fig 4.

In summary, our proposed hybrid structure containing elliptically-faceted (ELFA) diamond nanowire integrated with optical nanowire demonstrates a maximum unidirectional coupling efficiency of ~57 % for y-polarized dipole emissions. Further, and addition of BG provides a unidirectional coupling with maximum coupling efficiency of ~90 %. This structure can provide control over unidirectional coupling efficiency by optimizing the geometry of the ELFA diamond structure. We have exploited the core nano-domain interferometry for our structure and one can analyze this effect for other structure also. The structure can be also be used as TM-polarized supermode filter. Further, the structure can be implemented in various application in the fields of quantum optics, quantum memories, magnetometry, quantum computing, opto-spin based chemical sensing, and single-photon emission.

Acknowledgement: RJ acknowledge the support of SERB-STAR Fellowship, Govt. of India. This work is supported by STR/2020/000069.

Disclosures. The authors declare no conflicts of interest.

# Asymmetric Diamond Emitters for Unidirectional Photon Coupling: supplementary document

SATYAJIT MURMU, AVIJIT KUMAR, RAJAN JHA*

Nanophotonics and Plasmonics Laboratory, School of Basic Sciences
Indian Institute of Technology Bhubaneswar 752050, India
*Corresponding author: *rjha@iitbbs.ac.in

**SECTION S1. Effective mode indices of ELFA diamond nanowire coupled to optical nanowire**

Fig. S1 (a) shows a three-dimensional structure of elliptical facet (ELFA) diamond nanowire coupled to optical nanowire. The four rectangles represent four planes where the cross sections of the structure are of interest. Figure S1 (b) shows effective mode indices of TM-like supermode (y-polarized) along the z-direction in the elliptical-facet region. The insets show electric field intensity distribution of the first order supermode at the four different cross sections as in Fig. S1 (a). As the value of $L_{ex}$ increases, the refractive index of the first order supermode finally converges to the fundamental optical nanowire mode, but the second order supermode vanishes well before reaching the end of the elliptical facet [2].

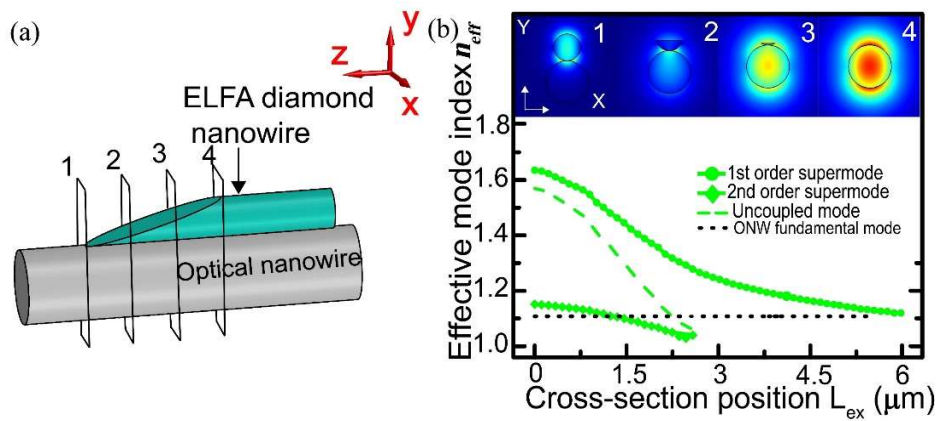

Fig. S1. (a) The proposed Hybrid ELFA structure (b) Effective mode indices of different cross-sections along elliptically region. The insets show the field intensity profiles of four first order supermode at the planes shown in panel (a). [adapted from ref. 2]

## SECTION S2. Electric field intensity at dipole plane as function of free length and dipole emissions for the ELFA structure coupled to optical nanowire

In Fig S2. (a), the reflection and transmission profile for various free length ($L_f$) values have been calculated. We find that the transmission profile depends on non-reflected two supermode powers at circular facet. On the other hand, the reflection profile depends on the first time reflected two supermode powers as well as second time reflected two supermode powers as shown in Fig. S2(b). Along with high periodicity modulations, we also observe spikes with periodicity of 60 nm. Fig S2. (c) shows electric field profile at the dipole plane (plane 4 in Fig. S1(a)) for $L_f$ values ranging from 10 nm to 330 nm. The field profile changes from first order supermode to second order supermode periodically with $L_f$ and that period is dependent on the phase difference between the two reflected and non-reflected same order supermode power. We also observe step-like feature is appearing the electric field profile at $L_f$ = 70, 130, 190, and 250 nm. At these $L_f$, the electric field is higher which results into enhanced Purcell factor.

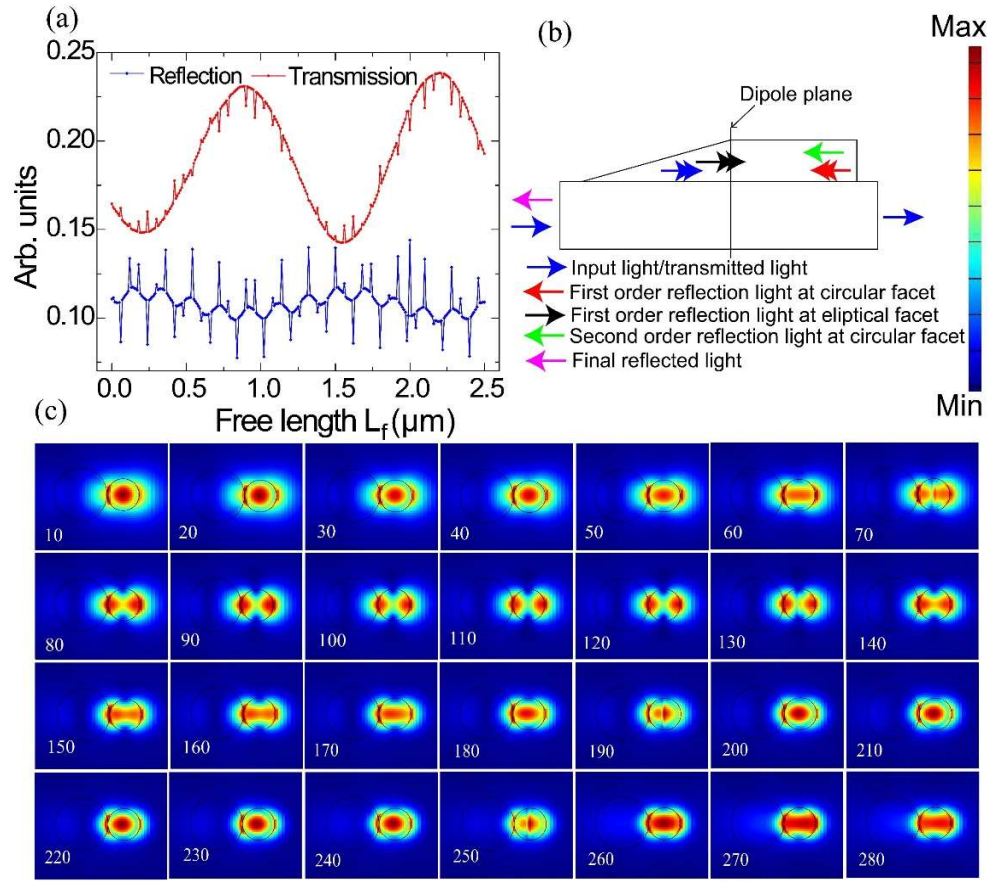

Fig. S2. (a) Transmission and reflection profile of hybrid Asymmetric ELFA structure for different $L_f$. (b) Various supermodes in the ELFA coupled to optical nanowire upon incident of a y-polarized light from the elliptical facet side. After light coupling to both the supermodes, some light will be reflected from the circular facet and propagates along elliptical facet direction in-order to couple in to optical nanowire fundamental mode. Transmitted part of power will couple to optical nanowire fundamental mode at circular facet. Again second time there will be reflection occurs in circular facet facet after light reflection in elliptical facet of ELFA diamond nanowire. Here double arrow represents first order and second order supermode and single arrow represent first order supermode. (c) Free length dependent electric field intensity plot at dipole plane for reflection spectra in ELFA structure.

## SECTION S3. Coupling efficiency and Purcell factor for the ELFA diamond nanowire coupled to optical nanowire

Fig.S3. represents $L_f$ dependent Purcell factor, $\Gamma$, and $C_{eff1}$ and $C_{eff2}$ for the ELFA structure. Here, we have chosen $L_e = 6$ µm corresponding to maximum unidirectional coupling efficiency as well as $\Gamma$. Figure Fig.S3. shows that $\Gamma$ exhibits an oscillatory behavior between 1.5 to 2.8 for $L_f$ larger than 4 µm. However, for $L_f < 4$ µm, it shows spikes of $L_f$. Further, the figure shows $L_f$ dependence of coupling efficiencies with $C_{eff1}$ behaving like simple oscillation w.r.t $L_f$ and $C_{eff2}$ shows an oscillatory behavior with additional larger period modulation. The oscillatory behavior of $C_{eff1}$ can be attributed to modal interference of reflected first order supermode from the circular facet of diamond nanowire and non-reflected first supermode power emanating from the dipole. Similarly, the reflected power of the two supermodes from the elliptical facet will also interfere with the two non-reflected supermode power directly from the emitter will determine $C_{eff2}$. The larger modulation period of $C_{eff2}$ is due to phase difference induced from different optical path length along free length region between supermode power and the smaller period of oscillation is due to phase variation of reflected light of same order supermode. The reflectance due to the circular facet is given by $((n_{eff}^i - n_f))/((n_{eff}^i + n_f))$, where $n_{eff}^i$ is ith order supermode index and $n_f$ is index of optical nanofiber. The periodicity is due to double optical path length of reflected second-order supermodes. The higher $L_e$ corresponds to distribution of higher first-order supermodes intensity and lower second-order supermodes. For $C_{eff2}$ this is reversed and behavior is different as shown in Fig. S3. i.e. it oscillates out phase with $C_{eff1}$. The behavior of $C_{eff2}$ is due to relatively prominent power in reflected and non-reflected second order supermodes. Here the phase variation is dependent of single optical length in free length region from reflected light and along with there are continuous phase variation between supermodes occurs from both reflected and non-reflected light emissions. For $C_{eff1}$ the phase variation is observed for only reflected light from circular facet.

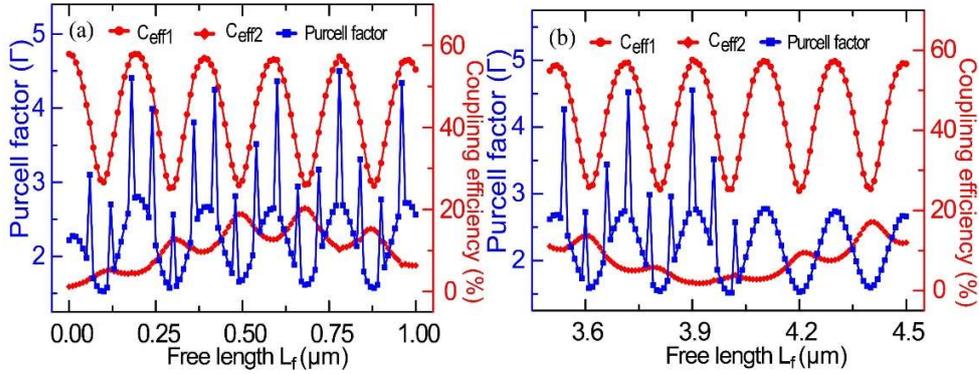

Fig. S3. (a)-(b) Free length dependent $\Gamma$, $C_{eff1}$, and $C_{eff2}$ of hybrid Asymmetric ELFA structure for fixed other structure parameter define as Fig. S2 and Fig. 3 (main text).

## SECTION S4. Free length dependence of coupling efficiencies $C_{eff1}$ and $C_{eff2}$ for the ELFA hybrid structure

In ELFA hybrid structure, the dipole emission couples to both the first ($s_1$) and the second order ($j_1$) supermodes towards to elliptically facet direction. Similarly, along circular facet direction the dipole emission also couples to both the first and the second order supermode power. Further at both facets lights in the both the supermodes go through reflection albeit with different reflection coefficient. $s_2$, $j_2$ are reflected first order, second order supermode power from the circular facet. The coupling of light in elliptical facet direction is the results of modal interference of two supermode power directly from the dipole emission (non-reflected) and two reflected supermodes power from circular facet. Similarly the coupling of light in circular facet is the result of modal interference of direct two supermode power (non-reflected) and two reflected supermode power from elliptical facet. We have used the Mach-Zehnder interference relation to calculate $C_{eff1}$ and $C_{eff2}$ for our hybrid ELFA structure as shown in Fig. S3. to understand their dependence on $L_f$. We find that the simple oscillation of $C_{eff1}$ is due the higher value of reflected second order supermode power from circular facet and negligible non-reflected second order supermode power (directly from the dipole). In the case of $C_{eff2}$ is it is reverse. The reflected second order and first order supermode power from elliptical facet is very less than non-reflected second order supermode power(directly from the dipole). The total power towards elliptical facet direction can be given by this equation inline with Mach-Zehnder interference[1].

$$S = s_1 + s_2 + j_1 + j_2 + 2\sqrt{(s_1\ s_2)}\cos(\varphi_1) + 2\sqrt{(j_1\ j_2)}\cos(\varphi_2) + \\ 2\sqrt{(s_1\ j_2)}\cos(\varphi_3) + 2\sqrt{(j_1\ s_2)}\cos(\varphi_4) + 2\sqrt{(s_1\ j_1)}\cos(\varphi_5) + \\ 2\sqrt{(s_2\ j_2)}\cos(\varphi_6) \qquad (1)$$

Here, $\varphi_1, \varphi_2, \varphi_3, \varphi_4, \varphi_5,$ and $\varphi_6$ are the phase differences between $_1$ and $s_2$; $j_1$ and $j_2$; $s_1$ and $j_2$; $j_1$ and $s_2$; $s_1$ and $j_1$; and $s_2$ and $j_2$. For various supermodes the phase difference ($\varphi_3, \varphi_4, \varphi_5,$ and $\varphi_6$) are given by

$$\varphi_n = \frac{2\pi L}{\beta_n}$$

Where $L$ is optical length, $\beta_n$ is beat length and can be obtained as below

$$\boldsymbol{\beta_n = \lambda/(n_{eff}^1 - n_{eff}^2)}.$$

The beat length for same supermode order ($\varphi_1,$ and $\varphi_2$) are given by

$$\boldsymbol{\beta_n = \lambda/n_{eff}^i}.$$

Where $i = 1,2$ for two supermode effective mode index in free length plane cross-section.

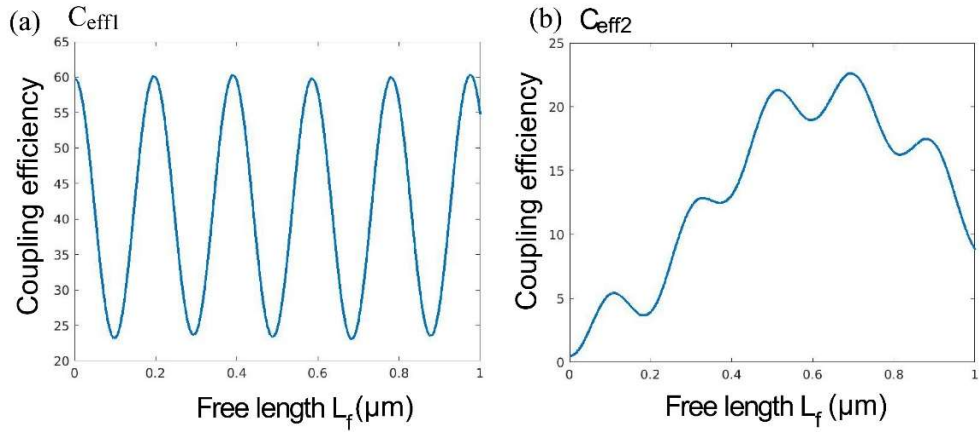

Fig. S4. Shows free length dependent (a) $C_{eff1}$ and (b) $C_{eff2}$ calculated using the total power from equation (1). Here, we have used effective mode indices from Fig. S1(b). Evidently $s_1$ and $s_2$ contribute significantly to $C_{eff1}$ while $j_1$ and $j_2$ contribute significantly to $C_{eff2}$.

**S5. Schematic Ray diagram of dipole emission coupling to various supermodes in ELFA structure**

Further a visualization of supermodes propagating within the structure can be given by the graphical representation as shown in Fig. S5.

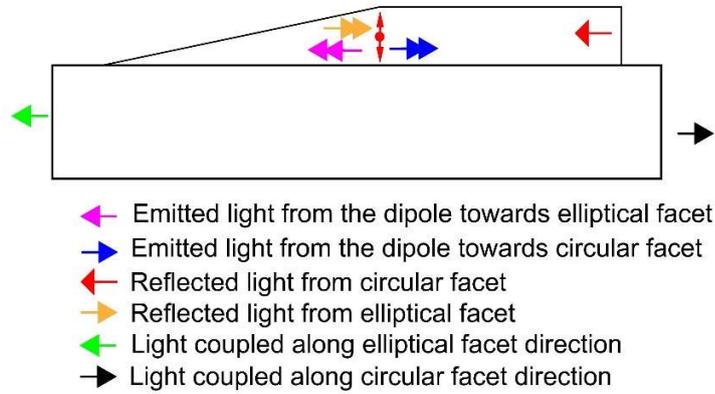

Fig. S5. Schematic diagram of dipole emission coupling to various supermodes and their propagation along both directions (elliptical facet and circular facet direction) and finally coupling to optical nanowire. A double-headed arrow indicates both the first and second order supermode powers and single-headed arrow indicates only fundamental mode.

**SECTION S6. Chirality constant's dependence on etched length**

The chirality constant for both ELFA and ELFA+BG structure has been shown in Fig. S6. We observe significantly higher chirality constant for ELFA+BG structure than ELFA hybrid structure of similar geometrical parameter indicating efficient unidirectional photon coupling.

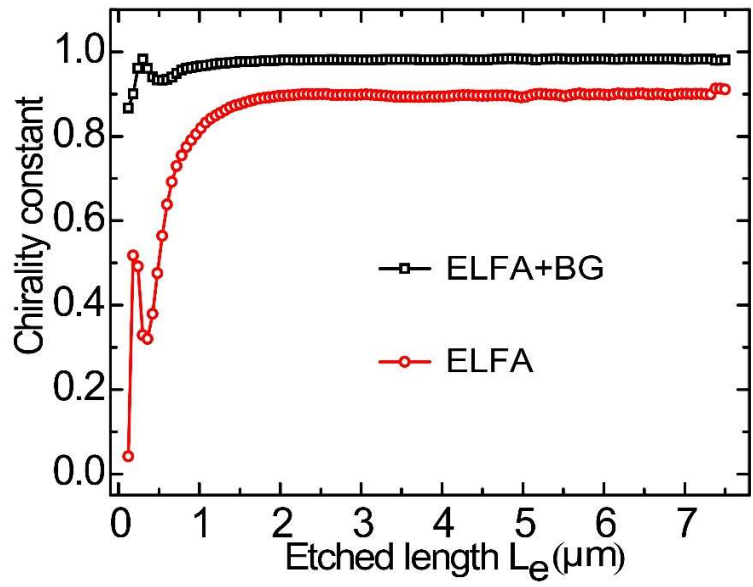

Fig. S6. shows chirality constant for both ELFA and ELFA+BG structure for various etched length $L_e$. Here the $L_f$ is 2.7 μm, $D_{nf}$ = 350 nm, and $D_{nd}$ = 220 nm.